\documentclass[a4paper,fleqn,usenatbib]{mnras}

\usepackage{newtxtext,newtxmath}
\usepackage[T1]{fontenc}
\usepackage{ae,aecompl}

\usepackage{graphicx}	% Including figure files
\usepackage{amsmath}	% Advanced maths commands
\usepackage{amssymb}	% Extra maths symbols

\title[Reflection spectra of thick disks]{Reflection spectra of thick accretion disks}

\author[S.~Riaz et al.]{
Shafqat~Riaz,$^{1}$
Dimitry~Ayzenberg,$^{1}$
Cosimo~Bambi$^{1}$\thanks{Corresponding author: bambi@fudan.edu.cn}
and Sourabh~Nampalliwar$^{2}$
\\
$^{1}$Center for Field Theory and Particle Physics and Department of Physics, Fudan University, 200438 Shanghai, China\\
$^{2}$Theoretical Astrophysics, Eberhard-Karls Universit\"at T\"ubingen, 72076 T\"ubingen, Germany
}

\pubyear{2019}

\begin{document}
\label{firstpage}
\pagerange{\pageref{firstpage}--\pageref{lastpage}}
\maketitle

\begin{abstract}
Relativistic reflection features are commonly observed in the X-ray spectra of stellar-mass and supermassive black holes and originate from illumination of the inner part of the accretion disk by a hot corona. All the available relativistic reflection models assume that the disk is infinitesimally thin and the inner edge is at the innermost stable circular orbit or at a larger radius. However, we know that several sources, especially among supermassive black holes, have quite high mass accretion rates. In such a case, the accretion disk becomes geometrically thick and the inner edge of the disk is expected to be inside the innermost stable circular orbit. In this work, we employ the Polish donut model to describe geometrically thick disks and we study the iron line shapes from similar systems. We also simulate full reflection spectra and we analyze the simulated observations with a thin disk relativistic reflection model to determine the impact of the disk structure on the estimation of the model parameters, in particular in the case of tests of the Kerr hypothesis.
\end{abstract}

\begin{keywords}
accretion, accretion discs -- black hole physics
\end{keywords}

%%%%%%%%%%%%%%%%%%%%%%%%%%%%%%%%%%%%%%%%%%%%%%%%%%

%%%%%%%%%%%%%%%%% BODY OF PAPER %%%%%%%%%%%%%%%%%%

\section{Introduction}

X-ray spectra of accreting black holes often show relativistic reflection features that are thought to be produced by illumination of the inner part of the accretion disk by a hot corona~\citep{i1,i2,i3,i4,i5,w13,i6,i7}. Thermal photons from the accretion disk can inverse Compton scatter off free electrons in the corona, which is some hot, usually compact and optically thin, cloud in the strong gravity region near the black hole (e.g. the accretion flow between the disk and the black hole, the atmosphere above the disk, the base of the jet, etc.). A fraction of the Comptonized photons illuminates the accretion disk, generating the reflection component. The latter is characterized by some narrow fluorescent emission lines in the soft X-ray band, in particular the iron K$\alpha$ complex at 6.4 to 6.97~keV depending on the ionization of iron atoms, and by the Compton hump peaked at 20-30~keV. In the presence of high quality data and with the correct astrophysical model, X-ray reflection spectroscopy can be quite a powerful tool to study the strong gravity region of black holes~\citep{t1,t2} and even to test Einstein's theory of general relativity~\citep{t3,t4,t5}.

Theoretical models to fit these reflection components are typically the combination of a non-relativistic reflection model calculating the reflection spectrum in the rest-frame of the gas, e.g. {\sc reflionx}~\citep{rl} or {\sc xillver}~\citep{x1,x2}, and a convolution model taking into account the structure of the accretion disk and the relativistic effects of the spacetime, e.g. {\sc kerrconv}~\citep{kc} or {\sc relconv}~\citep{rc1,rc2}. All the current convolution models employed to fit X-ray data of accreting black holes assume that the disk is infinitesimally thin, perpendicular to the black hole spin, and that the inner edge is at the innermost stable circular orbit (ISCO) or at a larger radius. However, in reality accretion disks around black holes have a finite thickness, which should increase with the mass accretion rate. The inner edge of the disk is thought to be at the ISCO for sources in the high-soft state with an accretion luminosity between 5\% to about 30\% of their Eddington limit~\citep{isco1,isco2}. For higher mass accretion rate, the gas pressure becomes more and more important, and the disk moves from thin to slim and then thick. The thickness of the disk increases, so the approximation of an infinitesimally thin disk gets worse and worse, and the inner edge of the disk may move inside the ISCO.

While the available relativistic reflection models should only be used to fit sources accreting from thin disks, they are commonly employed to fit any relativistic reflection spectrum~\citep{t1}. In particular in the case of supermassive black holes, we know that some sources have accretion luminosities well beyond 30\% of their Eddington limit, with some sources even accreting at the Eddington rate and above. This clearly has an impact on the estimation of the model parameters of the system. For example, this may lead to overestimating the black hole spin, because the inner edge of the disk is likely inside the ISCO and the reflection spectrum from the inner part of the disk may look like that of a disk with an inner edge at the ISCO in a spacetime with a smaller ISCO radius (and thus higher black hole spin). The issue is even more subtle if we want to analyze the reflection spectrum of a source to test the Kerr nature of the compact object~\citep{yerong,a0,a1}, because the employment of the wrong model may induce apparent deviations from the Kerr metric~\citep{yuexin,honghui}.

The aim of the present work is to calculate reflection spectra of thick disks and estimate the impact of the disk structure on the measurement of the parameters of the system when we fit the data with a thin disk model. In particular, we are interested in the impact on the estimation of the deformation parameters of the spacetime when we want to test the Kerr metric. Surprisingly, there is not much on this subject in the literature, despite it being quite clear that this effect exists and can lead to some erroneous estimates of the properties of accreting black holes. \citet{ww07} present iron line profiles for thick and sub-Keplerian accretion disks with inner edges at the ISCO or at larger radii. \citet{tr18} study the reflection spectrum from thin disks with inner edges at the ISCO radius for a disk model with finite thickness. In our study, we describe the accretion disk with the Polish donut model~\citep{kja78,abramowicz1978}, which is expected to be suitable for objects with very high accretion rates, say around the Eddington limit. Here the thickness of the disk is $H/R \sim 1$, where $H$ is the semi-height of the disk at the radius $R$. The inner edge of the disk is always inside the ISCO and can approach the marginally bound orbit.

We find that single line shapes of thick disks are definitively different from those generated by thin disks. However, when we consider the full reflection spectrum and we do not know the input values of the model parameters, it is very challenging to recover the correct values of the model parameters from the fit. In other words, some parameters show strong degeneracy, and the reflection spectrum of a thick disk can look like that of a thin disk with different values of some model parameters. Since the inner edge of the disk is inside the ISCO radius in the case of Polish donut disks, the spin parameter is usually overestimated. In the case of tests of the Kerr metric, a Kerr black hole may be interpreted as a non-Kerr compact object. However, as the spin parameter increases, real deviations from the Kerr background have quite a strong impact on the reflection spectrum and, when we fit the spectrum of a thick disk of a Kerr black hole with a thin disk model, we may not measure deviations from the Kerr geometry even if we still overestimate the black hole spin.

The content of the paper is as follows. In Section~\ref{s-disk}, we briefly review the Polish donut model, which is the model adopted in this work to describe thick accretion disks. In Section~\ref{s-lines}, we present our convolution model and we show iron line shapes expected from thick disks described by the Polish donut model for different values of the black hole spin and of the inclination angle of the disk. In Section~\ref{s-sim}, we consider the whole reflection spectrum and we simulate some observations with \textsl{NICER}~\citep{nicer} to determine the impact of the disk structure on the estimation of the black hole spin and the deformation parameter of the spacetime. We discuss our results in Section~\ref{s-con}. In Appendix~\ref{a-j}, we review the Johannsen metric~\citep{tj}, which is the spacetime employed in the reflection model used in Section~\ref{s-sim} to analyze the simulations and estimate the impact of the disk structure on the measurement of the model parameters. Throughout the manuscript, we adopt units in which $c = G_{\rm N} = 1$ and a metric with signature $(-+++)$.

%%%%%%%%%%%%%%%%%%%%%%%%%%%%%%%%%%%%%%%%%%%%%%%%%%

\section{Accretion disk model}
\label{s-disk}

Analytic models for relativistic thick accretion disks with negligible self-gravitation were originally developed in \citet{FM1976}, \citet{kja78}, and \citet{abramowicz1978}. These authors considered stationary, axially symmetric, perfect fluid disks in generic stationary, axisymmetric, and asymptotically flat spacetimes. However, \citet{FM1976} assumed isentropic disks, while \citet{kja78} and \citet{abramowicz1978} worked with barotropic disks. The latter is today known as the ``Polish donut'' model and describes a low viscosity, geometrically and optically thick, radiation pressure supported disk undergoing super-Eddington black hole accretion. In this section, we just provide a brief overview. More details can be found, for instance, in \citet{zilong} and \citet{book}.

We assume a stationary, axisymmetric, and asymptotically flat spacetime. We write the line element as\footnote{We note that the most general stationary, axisymmetric, and asymptotically flat spacetime also has a non-vanishing $g_{tr}$ [see, for instance, Section~7.1 in~\citet{wald}], but here we ignore such a possibility. In general relativity in vacuum, we can always find a coordinate system with the line element as in Eq.~(\ref{line_element_PD}) where $g_{tr} = 0$.}
\begin{eqnarray}
\label{line_element_PD}
ds^2 = g_{tt}dt^2 + g_{rr}dr^2+g_{\theta\theta}d\theta^2 + 2g_{t\phi}dtd\phi + g_{\phi\phi}d\phi^2 \, ,
\end{eqnarray}
where the metric elements are all independent of the $t$ and $\phi$ coordinates. The matter in the disk is treated as a perfect fluid with purely azimuthal flow. The  stress-energy tensor of the system can be written as
\begin{eqnarray}
\label{stress_energy_tensor}
T^{\mu\nu} = (\rho + P)u^{\mu}u^{\nu} + g^{\mu\nu}P \, ,
\end{eqnarray}   
where $\rho$ and $P$ are, respectively, the energy density and the pressure of the perfect fluid, which are related through a barotropic equation of state $\rho = P(\rho)$. The 4-velocity vector of the fluid element is 
\begin{eqnarray}
\label{four_velocity}
u^{\mu} = (u^{t}, 0, 0, u^{\phi}) \, .
\end{eqnarray}   
The specific energy of the fluid element is $-u_t$. The angular velocity and the angular momentum per unit energy are $\Omega = u^{\phi}/u^{t}$ and $l = -u_{\phi}/u_{t}$, respectively. $u_t$, $\Omega$, and $l$ can be written in terms of metric elements as
\begin{eqnarray}
\label{ut}
u_{t} &=& -\sqrt{\frac{g_{t\phi}^2 -g_{tt}g_{\phi\phi} }{g_{\phi\phi} + 2lg_{t\phi} + l^2g_{tt}}} \, , \\
\label{omega}
\Omega &=& \frac{lg_{tt} + g_{t\phi}}{lg_{t\phi} + g_{\phi\phi}} \, , \\
\label{angular momentum}
l &=& - \frac{g_{t\phi} + \Omega g_{\phi\phi}}{g_{tt} + \Omega g_{t\phi}} \, .
\end{eqnarray}
The motion of a fluid element in the disk is governed by Euler's equation 
\begin{eqnarray}
\label{euler equation}
\nabla_{\nu} T^{\mu\nu} = 0 \, ,
\end{eqnarray}  
which leads to the following equation for the 4-acceleration of the fluid element $a^{\mu} = u^{\nu}\nabla_{\nu}u^{\mu}$
\begin{eqnarray}
\label{4-accleration}
a^{\mu} = - \frac{g^{\mu\nu} + u^{\mu}u^{\nu}}{\rho + P}\partial_{\nu} P \, . 
\end{eqnarray}

Exploiting the assumptions that the spacetime is stationary and axisymmetric and that the equation of state is barotropic, the 4-acceleration of the fluid element can be written as a gradient of the scalar potential $W(P)$
\begin{eqnarray}
\label{4-acceleration as gradient of scalar potential}
a_{\mu} = \partial_{\mu}W \, , \quad  W(P) = -\int^P \frac{dP^{\prime}}{\rho(P^{\prime}) + P^{\prime}} \, .
\end{eqnarray}    
From Eq.~(\ref{euler equation}) we can infer that there exists an invariant function $\Omega = \Omega(l)$~\citep{abramowicz1978}. Integrating Eq.~(\ref{euler equation}), we find
\begin{eqnarray}
\label{potential}
W = W_{\rm in} + ln\frac{u_{t}}{u_{t}^{\rm in}} + \int_{l_{\rm in}}^{l} \frac{\Omega d l^\prime}{1-\Omega l^{\prime}} \, ,
\end{eqnarray}    
where $W_{\rm in}$, $-u_{t}^{\rm in}$ and $l_{\rm in}$ are, respectively, the potential, the energy per unit mass, and the angular momentum per unit energy at the inner edge of the disk and can be replaced by the same quantities at any other point of the fluid boundary. In the Newtonian limit, $W$ reduces to the sum of the gravitational and the centrifugal potential and vanishes at infinity. For the special case $l = {\rm constant}$, Eq.~(\ref{potential}) still holds, and the integral vanishes.

Once the background metric is known, the function $\Omega = \Omega(l)$ determines the dissipative process and characterizes the fluid rotation in the disk as well. In the zero viscosity case, this function cannot be determined from any equation and it should come from the assumption on the model. If we fix $\Omega = \Omega(l)$, we can determine the surfaces of the constant pressure of the fluid, which are usually called the equipotential surfaces and have $W = {\rm constant} < 0$. Such surfaces describe possible boundaries of the fluid. Accretion is possible when the fluid surface has one (or more) cusp(s): the fluid fills out the Roche lobe and is then transferred to the compact object~\citep{abramowicz1978,zilong,book}. The mechanism does not need any fluid viscosity, which, at least in principle, could thus be very low.

For simplicity, in this work we consider the special case $l = {\rm constant}$, which is marginally stable with respect to axisymmetric perturbations. For $l = {\rm constant}$, Eq.~(\ref{potential}) reduces to 
\begin{eqnarray}
\label{reduced effective potential 10}
W = \ln(-u_{t}) + {\rm constant} \, .
\end{eqnarray} 
In the Kerr spacetime, there are five qualitatively different configurations, depending upon the value of $l$. The most interesting and relevant to the present work is the one in which $l_{\rm ms}<l<l_{\rm mb}$, where $l_{\rm mb}$ and $l_{\rm ms}$ are, respectively, the angular momentum per unit energy at the marginally bound and the marginally stable (or ISCO) orbits. For these values of $l$, there is only one surface with cusp $(W = W_{\rm cusp})$ and the location of the cusp lies in between the marginally stable and marginally bound orbits. The configurations with surface $W < W_{\rm cusp}$ describe non-accreting disks, while those with surface $W > W_{\rm cusp}$ do not represent any disk configuration~\citep{abramowicz1978,zilong,book}.

%%%%%%%%%%%%%%%%%%%%%%%%%%%%%%%%%%%%%%%%%%%%%%%%%%

\section{Iron line shapes}
\label{s-lines}

Let us consider a black hole accreting from a geometrically and optically thick disk, which we will describe with the Polish donut model. The illumination of the disk by the corona produces a reflection component. In order to understand the impact of the disk structure on the reflection spectrum, in this section we will simplify the whole reflection spectrum with a single iron line at 6.4~keV. Full reflection spectra will be considered in the next section.

Calculations of single iron lines of thin accretion disks have already been described in the literature; see, e.g., \citet{t3} or \citet{book}. We consider a distant observer with viewing angle $i$ and we calculate backward in time photon trajectories from the image plane of the distant observer to the emission points on the accretion disk. When a photon hits the disk, we calculate the redshift factor 
\begin{eqnarray}
\label{g_factor}
g = \frac{\sqrt{-g_{tt} - 2g_{t\phi}\Omega - g_{\phi\phi}\Omega^2}}{1 + \lambda\Omega} \, , 
\end{eqnarray}        
where $\Omega$ is the angular velocity of the fluid element, $\lambda = k_{\phi}/k_{t}$, and $k_t$ and $k_\phi$ are, respectively, the $t$ and $\phi$ components of the photon 4-momentum. In a stationary and axisymmetric spacetime, $k_t$ and $k_\phi$ are constants of motion, and therefore $\lambda$ is a constant of motion too and can be evaluated from the photon initial conditions. Finally, the iron line shape of the accretion disk is obtained by integrating over the disk image
\begin{eqnarray}
\label{photon_flux_liouville_theorem}
N(E_{\rm obs}) = \frac{1}{E_{\rm obs}}\int g^3 I_{\rm e} (E_{\rm e})\frac{dX dY}{D^2} \, ,
\end{eqnarray}
where $N(E_{\rm obs})$ is the photon count with energy $E_{\rm obs}$ (as measured by the distant observer), $I_{\rm e}$ and $E_{\rm e}$ are, respectively, the specific intensity of radiation and the photon energy at the emission point on the disk (as measured in the rest frame of the accreting gas), $X$ and $Y$ are the Cartesian coordinates of the image plane of the distant observer, and $D$ is the distance of the source from the distant observer. In the case of monochromatic emission (single line) with power-law profile, the specific intensity of radiation $I_{\rm e}$ can be written as 
\begin{eqnarray}
\label{emitted_intensity}
I_{\rm e}(E_{\rm e}) \propto \frac{\delta (E_{\rm e} - E_{\rm line})}{r^q} \, ,
\end{eqnarray}       
where $E_{\rm line} = 6.4$~keV for a neutral iron K$\alpha$ line and $q$ is the emissivity index. More details can be found, for instance, in \citet{t3} or \citet{book}.

In the case of a thick disk described by the Polish donut model, the calculations are quite similar. We still start from the image plane of the distant observer and we calculate backward in time the null geodesics of a grid of photons to obtain the image map of the accretion disk, with photon redshift and photon intensity at every pixel of the image. For a specific metric, we fix the value of $l$ and we infer the boundary of the accretion disk. Photon trajectories are still calculated until the photon hits the disk, but now the emission point is outside the equatorial plane and we need a more sophisticated routine to determine its coordinates. Once the coordinates of the emission point are determined, we still employ Eq.~(\ref{g_factor}) to infer the redshift factor. Now $\Omega$ is calculated from Eq.~(\ref{omega}), while in the thin disk case $\Omega$ was the Keplerian angular velocity. Note also that some parts of the disk may not be observed as they are obscured by the disk itself; the phenomenon does not occur for face-on disks, becomes more and more important as the viewing angle increases, and the inner part of the disk is completely obscured for high viewing angles.

\begin{figure}
\begin{center}
\includegraphics[width=8.5cm]{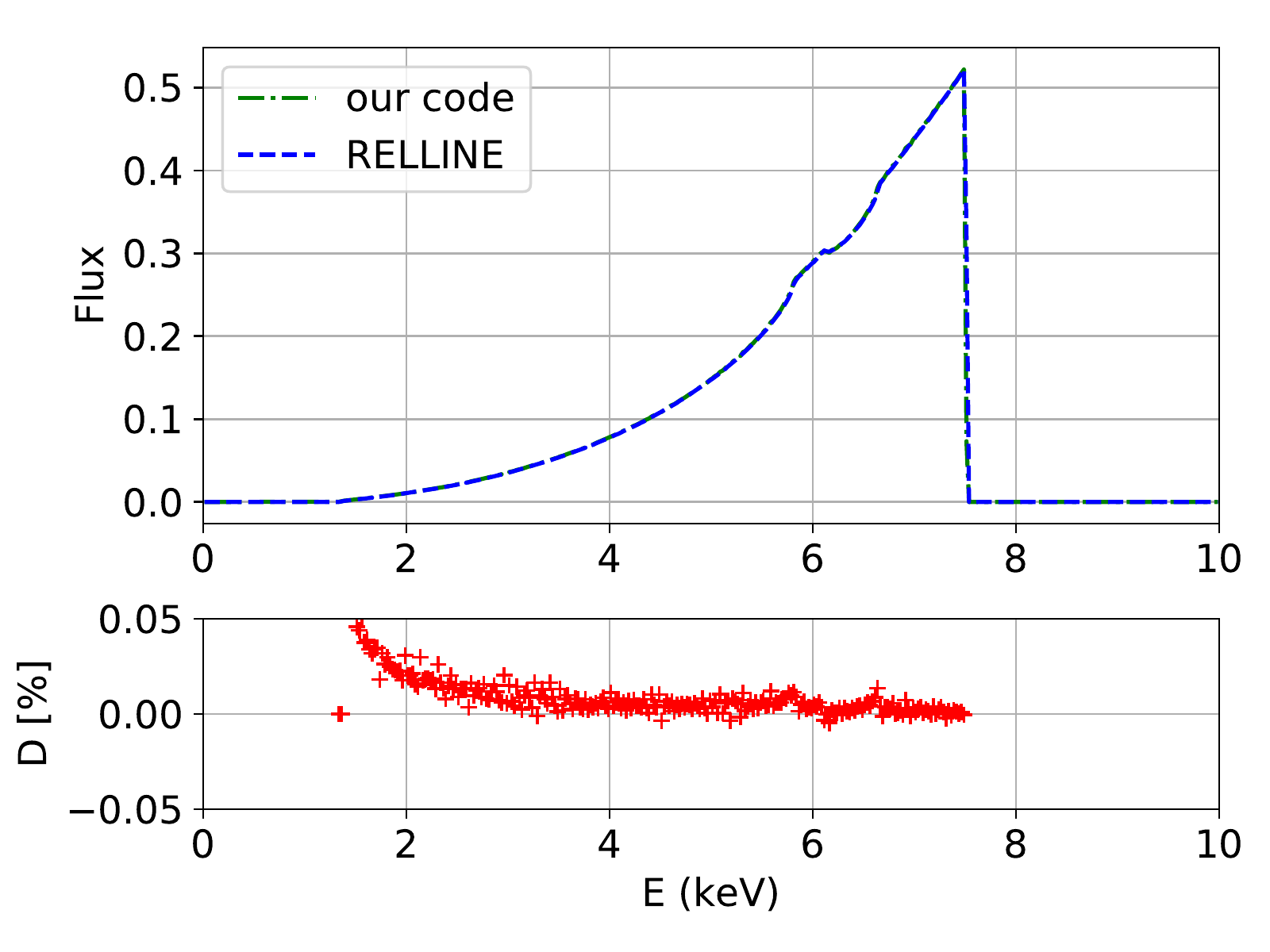}
\vspace{-0.3cm}
\caption{Comparison of a single iron line as calculated with our ray-tracing code and {\sc relline} in the case of a infinitesimally thin disk in the Kerr spacetime with spin parameter $a_{*} = 0.9$, viewing angle $i = 55^{\circ}$, and emissivity index $q = 3$. The inner edge of the disk is set at the ISCO and the outer edge of the disk is at $R_{\rm out} = 400$~$M$. Top panel: photon flux (in arbitrary units) as a function of the photon energy (in keV and assuming $E_{\rm line} = 6.4$~keV). Bottom panel: relative difference between the photon fluxes calculated by our code and by {\sc relline}.}
\label{f-raytracing}
\end{center}
%\end{figure}
\vspace{-0.4cm}
%\begin{figure}
\begin{center}
\includegraphics[width=8.5cm]{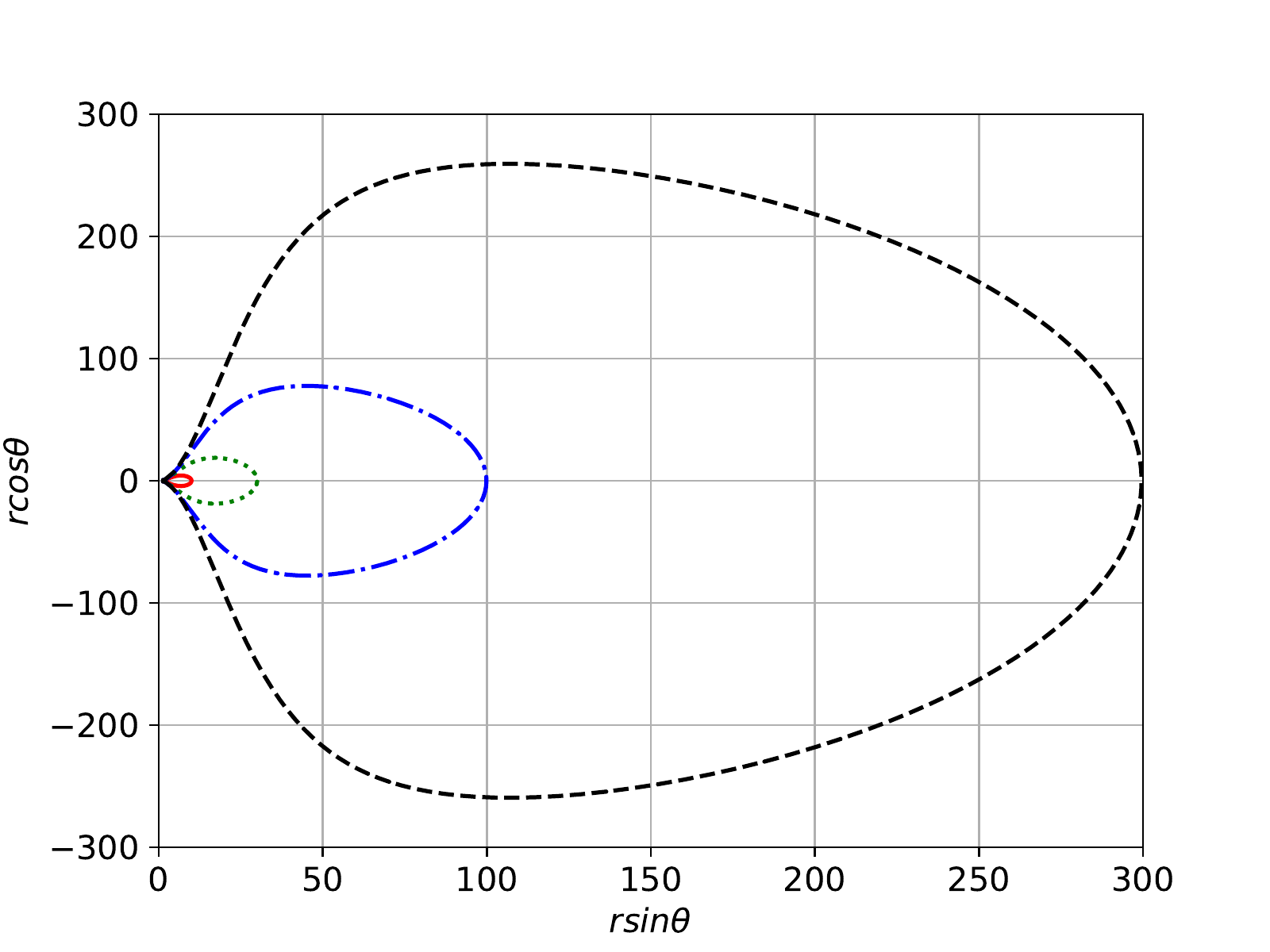}
\vspace{-0.3cm}
\caption{Examples of Polish donut disks in Kerr spacetime with spin parameter $a_* = 0.95$ for different values of the outer edge $R_{\rm out}$: $R_{\rm out} = 10$~$M$ (red solid curve), $30$~$M$ (green dotted curve), $100$~$M$ (blue dotted-dashed curve), and $300$~$M$ (black dashed curve). $r \sin\theta$ and $r \cos\theta$ in units $M = 1$.}
\label{f-donut}
\end{center}
\end{figure}

The ray-tracing code to calculate the photon trajectories from the image plane of the distant observer to the emission point on the accretion disk is that described in~\citet{noi2}. However, unlike the code in~\citet{noi2} that is used to calculate the transfer function of the spacetime, we directly evaluate the photon count per energy bin using Eq.~(\ref{emitted_intensity}), as done in \citet{cfm}. In such a case, it is easier to reach a better accuracy since we skip the numerical uncertainties related to the calculation of the transfer function, its tabulation, and its integration. We have compared our calculations for thin disks with {\sc relline}~\citep{rc1,rc2}, finding that the discrepancy in the photon flux is below 0.1\%, see Fig.~\ref{f-raytracing}. We cannot do the same for our calculations for thick disks because there are no existing models for a similar check.

\begin{figure*}
\vspace{0.3cm}
\includegraphics[width=17.5cm]{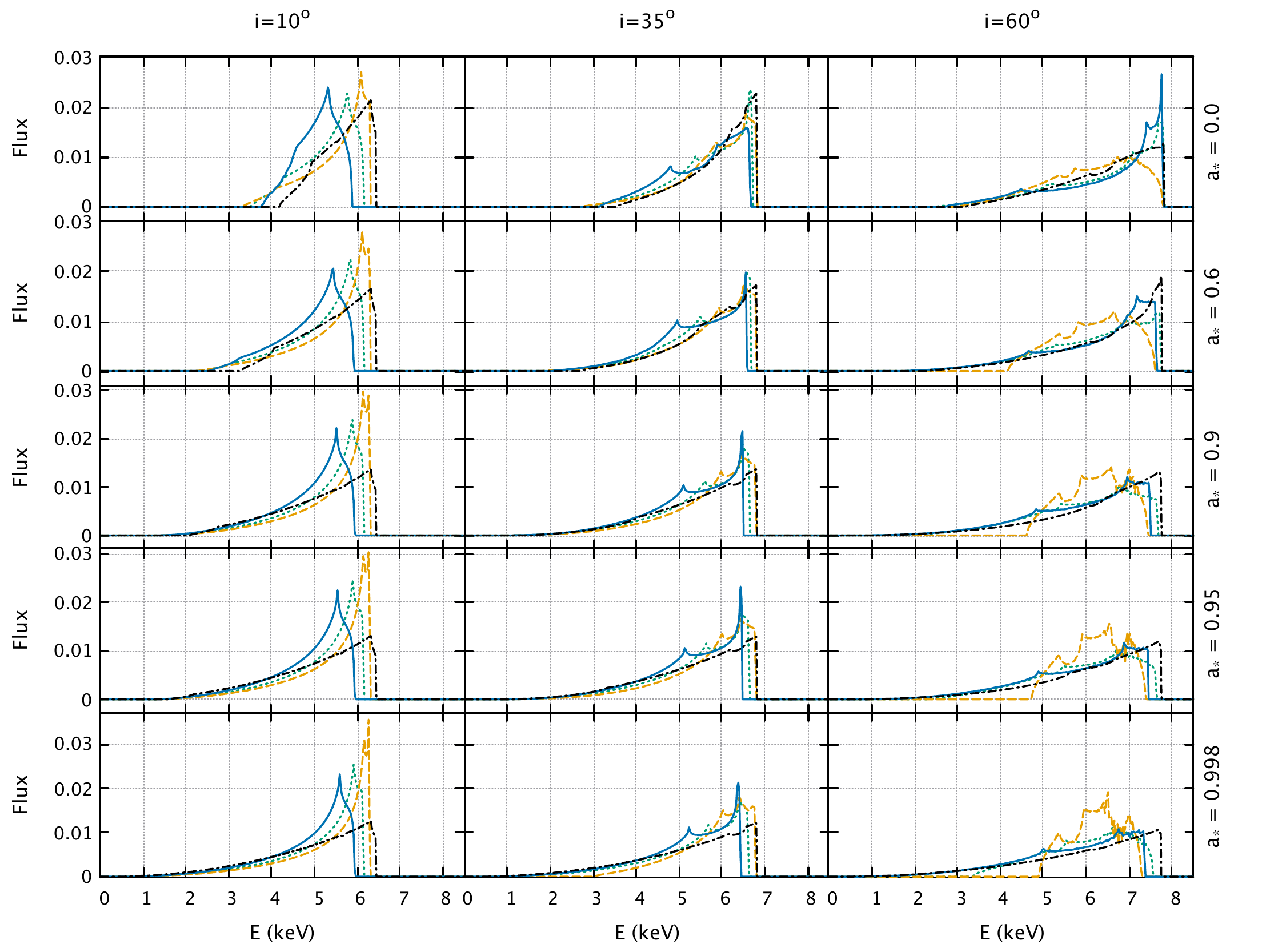}
\caption{Iron line shapes for different black hole spins and viewing angles for an infinitesimally thin Novikov-Thorne disk (black dotted-dashed curves) and Polish donut disks with outer radius, respectively, 12~$M$ (blue solid curves), 20~$M$ (green dotted curves), and 40~$M$ (orange dashed curves). The emissivity profile of the disk is modeled with a power-law with emissivity index $q = 3$. Flux in arbitrary units.}
\label{f-lines3}
\end{figure*}

\begin{figure*}
\vspace{0.3cm}
\includegraphics[width=17.5cm]{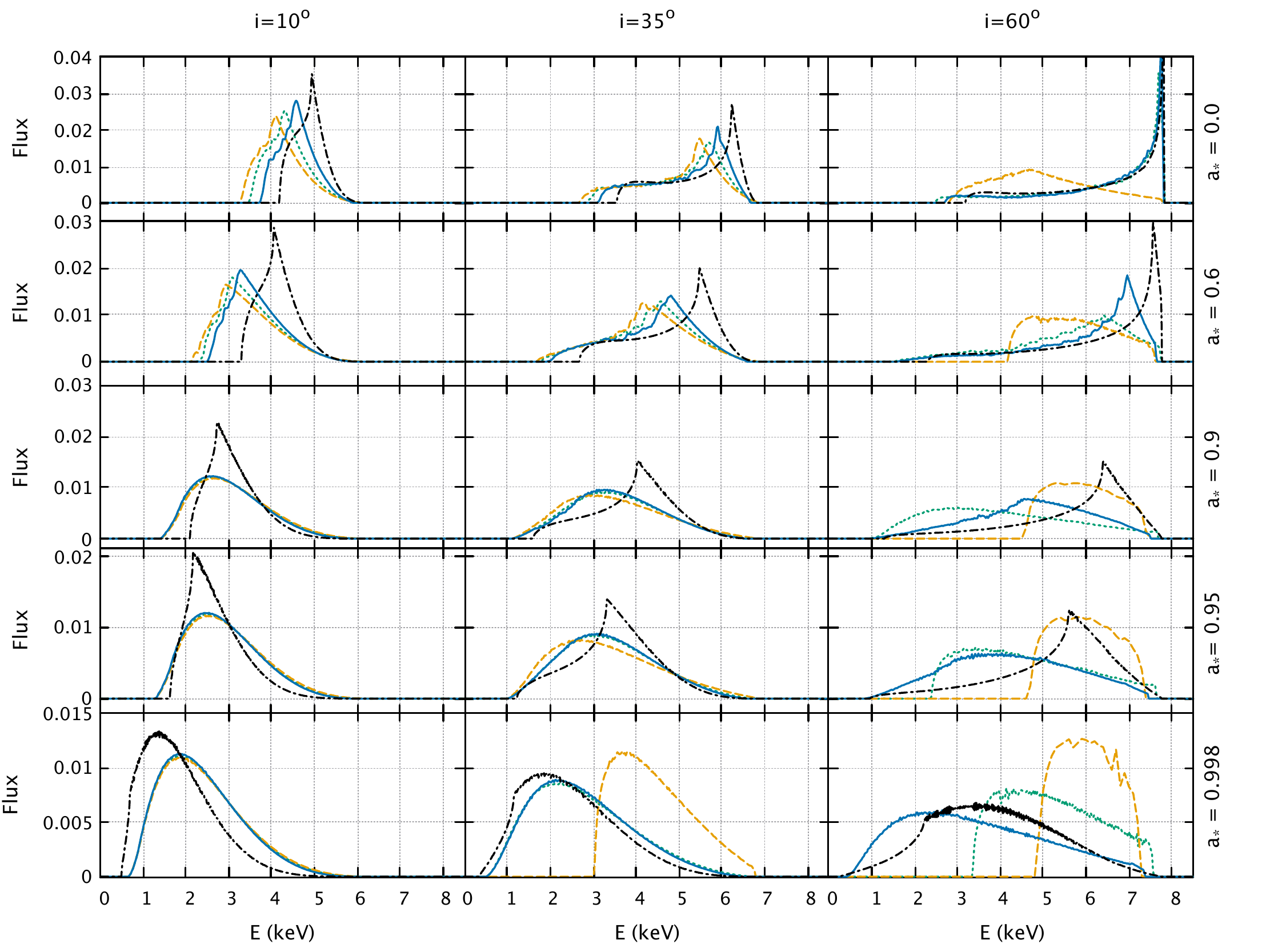}
\caption{As in Fig.~\ref{f-lines3} but the emissivity profile of the disk is now modeled with a power-law with emissivity index $q = 9$.}
\label{f-lines9}
\end{figure*}

Iron line shapes calculated by our ray-tracing code for $q = 3$ and 9 are shown, respectively, in Fig.~\ref{f-lines3} and Fig.~\ref{f-lines9} for a dimensionless spin parameter $a_* = 0$, 0.6, 0.9, 0.95, and 0.998 (from top to bottom) and for a viewing angle $i = 10^\circ$, $35^\circ$, and $60^\circ$ (from left to right). In every panel, the black dotted-dashed curve indicates the iron line of a thin disk with inner edge at the ISCO and outer edge $R_{\rm out} = 400$~$M$. For the Polish donut disks, we do not fix $l$ but the outer edge of the disk $R_{\rm out}$ in order to compare similar size disks in different spacetimes. The outer edge of the disk is at $R_{\rm out} = 12$~$M$ (blue solid curves), $20$~$M$ (green dotted curves), and $40$~$M$ (orange dashed curves). Such values of the outer edges are unnaturally low for real accretion disks, but, within the Polish donut model, larger disks do not permit the observation of their inner part except for a viewing angle $i$ close to $0^\circ$ (face-on disks), see Fig.~\ref{f-donut}. This fact challenges the comparison between our thin and thick disks because of the finite size effects of the thick disks, but larger thick disks would make the comparison completely impossible because their spectrum does not include the strongly redshifted photons from the region of the disk closer to the black hole. The choice of the high emissivity index $q = 9$ is to minimize such a problem related to the Polish donut model, as with such a high value of $q$ most of the radiation is emitted from the inner edge. The inner edge of the disk is always inside the ISCO. In Fig.~\ref{f-lines3} and Fig.~\ref{f-lines9} we see that the iron lines of the polish donut model have some bumpiness, particularly at high inclination. This also shows up in the thin disk lines at higher spin in Fig.~\ref{f-lines9}. This is just the result of some numerical uncertainty in the calculations, but it is completely negligible for what follows.

The iron line shapes of thick accretion disks are remarkably different from those of thin disks in the same spacetimes and with the same inclination angles. For $q=3$, the intensity of the radiation decays slower as the radial coordinate increases, so the effects of the outer edge of the disk in the Polish donut model are more pronounced. For $q = 9$, most of the radiation comes from the very inner part of the accretion disk, so the differences in the line shapes is mainly determined by disk structure near the inner edge. For thicker disks (larger $R_{\rm out}$) and/or higher viewing angles, the inner part of the disk may be partially or completely obscured, with the result that the low energy tail of the line (due to strongly redshifted photons emitted near the inner edge of the disk) is missing. This can be quite clearly seen from the orange curves for $i = 60^\circ$, both in Fig.~\ref{f-lines3} and Fig.~\ref{f-lines9}. If we increase the value of the spin parameter $a_*$, the inner edge of the disk moves to smaller radii. In general, this should produce highly redshifted photons. However, in the case of thick disks such photons may not reach the distant observer, except in the case of almost face-on disks, so a higher spin parameter does not necessarily imply a more redshifted iron line.

%%%%%%%%%%%%%%%%%%%%%%%%%%%%%%%%%%%%%%%%%%%%%%%%%%

\begin{table*}
\centering
\caption{Input parameters and best-fit values for simulations~A-D. The reported uncertainties correspond to 90\% confidence level for one relevant parameter. $^\star$ indicates that the parameter is frozen in the fit.}
\label{t-fit}
{\renewcommand{\arraystretch}{1.3}
\begin{tabular}{lccccccccccc}
\hline\hline
 & \multicolumn{2}{c}{Simulation~A} && \multicolumn{2}{c}{Simulation~B} && \multicolumn{2}{c}{Simulation~C} && \multicolumn{2}{c}{Simulation~D} \\
 & Input & Fit && Input & Fit && Input & Fit && Input & Fit \\
\hline
{\sc tbabs} &&&&&&&&&& \\
$N_{\rm H} / 10^{20}$ cm$^{-2}$ & $6.74$ & $6.74^\star$ && $6.74$ & $6.74^\star$ && $6.74$ & $6.74^\star$ && $6.74$ & $6.74^\star$ \\
\hline
{\sc relxill\_nk} &&&&&&&& \\
$q$ & $9$ & $4.979^{+0.011}_{-0.010}$ && $9$ & $5.306^{+0.011}_{-0.034}$ && $9$ & $5.34^{+0.04}_{-0.05}$ && $9$ & $8.28^{+0.09}_{-0.21}$ \\
$i$ [deg] & $35$ & $27.05^{+0.14}_{-0.18}$ && $35$ & $31.38^{+0.24}_{-0.14}$ && $35$ & $31.7^{+0.5}_{-0.3}$ && $35$ & $46.3^{+3.5}_{-0.4}$ \\
$a_*$ & $0.80$ & $0.9251^{+0.0010}_{-0.0011}$ && $0.85$ & $0.9727^{+0.0008}_{-0.0012}$ && $0.90$ & $0.9910^{+0.0009}_{-0.0011}$ && $0.95$ & $>0.997$ \\
$\alpha_{13}$ & $0$ & $-1.08^{+0.02}_{-0.02}$ && $0$ & $-0.694^{+0.026}_{-0.002}$ && $0$ & $-0.567^{+0.025}_{-0.010}$ && $0$ & $0.000^{+0.014}_{-0.013}$ \\
$\log\xi$ & $3.1$ & $3.0984^{+0.0008}_{-0.0006}$ && $3.1$ & $3.1074^{+0.0006}_{-0.0011}$ && $3.1$ & $3.1074^{+0.0013}_{-0.0013}$ && $3.1$ & $3.070^{+0.003}_{-0.003}$ \\
$A_{\rm Fe}$ & $1$ & $1.038^{+0.009}_{-0.010}$ && $1$ & $1.070^{+0.016}_{-0.009}$ && $1$ & $1.085^{+0.018}_{-0.021}$ && $1$ & $0.994^{+0.032}_{-0.015}$ \\
$\Gamma$ & $2$ & $2.0024^{+0.0003}_{-0.0004}$ && $2$ & $1.9965^{+0.0009}_{-0.0005}$ && $2$ & $1.9962^{+0.0011}_{-0.0005}$ && $2$ & $2.0008^{+0.0005}_{-0.0030}$ \\
$E_{\rm cut}$ [keV] & $300$ & $300^\star$ && $300$ & $300^\star$ && $300$ & $300^\star$ && $300$ & $300^\star$ \\
$R_{\rm f}$ & -- & $> 8$ && -- & $> 8$ && -- & $> 9$ && -- & $0.698^{+0.013}_{-0.042}$ \\
\hline
$\chi^2/\nu$ && $\quad 1185.88/1169 \quad$ &&& $\quad 1092.35/1169 \quad$ &&& $\quad 1107.52/1169 \quad$ &&& $\quad 1145.82/1169 \quad$ \\
&& = 1.01444 &&& = 0.934429 &&& = 0.947411 &&& = 0.980175 \\
\hline\hline
\end{tabular}}
\end{table*}

\begin{table*}
\centering
\caption{Input parameters and best-fit values for simulations E and F. The reported uncertainties correspond to 90\% confidence level for one relevant parameter. $^\star$ indicates that the parameter is frozen in the fit. $^\dag$ indicates that the parameter is unconstrained and we just report the best-fit value.}
\label{t-fit2}
{\renewcommand{\arraystretch}{1.3}
\begin{tabular}{lccccc}
\hline\hline
 & \multicolumn{2}{c}{Simulation~E} && \multicolumn{2}{c}{Simulation~F} \\
 & Input & Fit && Input & Fit \\
\hline
{\sc tbabs} &&&&& \\
$N_{\rm H} / 10^{20}$ cm$^{-2}$ & $6.74$ & $6.74^\star$ && $6.74$ & $6.74^\star$ \\
\hline
{\sc relxill\_nk} &&&&& \\
$q$ & $9$ & $5.136^{+0.025}_{-0.035}$ && $9$ & $3.847^{+0.006}_{-0.005}$ \\
$i$ [deg] & $35$ & $28.5^{+0.5}_{-2.0}$ && $35$ & $\sim 3^\dag$ \\
$a_*$ & $0.95$ & $0.9921^{+0.0005}_{-0.0012}$ && $0.95$ & $0.9131^{+0.0005}_{-0.0007}$ \\
$\alpha_{13}$ & $0.3$ & $-0.61^{+0.11}_{-0.06}$ && $-0.3$ & $-1.85^{+0.04}_{-0.03}$ \\
$\log\xi$ & $3.1$ & $3.0934^{+0.0016}_{-0.0019}$ && $3.1$ & $3.0973^{+0.0007}_{-0.0007}$ \\
$A_{\rm Fe}$ & $1$ & $1.016^{+0.024}_{-0.020}$ && $1$ & $1.130^{+0.010}_{-0.010}$ \\
$\Gamma$ & $2$ & $2.0072^{+0.0007}_{-0.0011}$ && $2$ & $2.0149^{+0.0003}_{-0.0003}$ \\
$E_{\rm cut}$ [keV] & $300$ & $300^\star$ && $300$ & $300^\star$ \\
$R_{\rm f}$ & -- & $> 8$ && -- & $4.58^{+0.04}_{-0.23}$ \\
\hline
$\chi^2/\nu$ && $\quad 1192.34/1169 \quad$ &&& $\quad 1447.23/1169 \quad$ \\
&& =1.01997 &&& =1.23801 \\
\hline\hline
\end{tabular}}
\end{table*}

\section{Reflection spectra, simulations, and impact on parameter estimates}
\label{s-sim}

All the available relativistic reflection models assume infinitesimally thin accretion disks with the inner edge at the ISCO or at a larger radius. In principle, such models should only be used to fit the reflection spectra of sources with thin disks. In reality, these models are commonly used to fit relativistic reflection features of any source, regardless of its mass accretion rate. This, of course, leads to systematic uncertainties in the final measurements of the model parameters. For instance, Tab.~1 in \citet{t1} shows a summary of black hole spin measurements derived from the analysis of reflection features and an estimate of the accretion luminosity of the corresponding source\footnote{We note that Tab.~1 in \citet{t1} shows the total luminosity of every source. A better proxy for the accretion luminosity and the disk thickness would be the disk luminosity.}. If the latter are correct, some accretion disks must be definitively thick.

In our case, we are particularly interested in the impact of the disk structure on tests of the Kerr metric~\citep{review}. In a certain sense, it is indeed surprising that it is possible to obtain very strong constraints on possible deviations from the Kerr metric using thin disk relativistic reflection models from supermassive black holes that should not accrete from thin disks, see \citet{a1}. Naively, we could expect that the systematic uncertainties of the model can lead to measurements of deviations from the Kerr spacetime. On the contrary, in \citet{a1}, we intentionally selected very-fast rotating black holes (with no preference based on accretion luminosity and some sources are indeed thought to exceed their Eddington limit) and we found very stringent constraints on possible deviations from the Kerr background. Here, we want to understand how this is possible by simulating some reflection spectra of thick disks and by fitting the faked data with our thin disk relativistic reflection model {\sc relxill\_nk} to test the Kerr hypothesis~\citep{noi1,noi2}\footnote{{\sc relxill\_nk} is an extension of the {\sc relxill} package~\citep{rc2,x2} to non-Kerr spacetimes and the two last letters NK stands for Non-Kerr. See~\cite{noi1,noi2} for more details.}.

To simulate reflection spectra, we use the ray-tracing code described in the previous section to calculate the photon flux via Eq.~(\ref{photon_flux_liouville_theorem}). Now the specific intensity of the radiation at the emission point, $I_{\rm e}$, is not given by Eq.~(\ref{emitted_intensity}) but is calculated with {\sc xillver}~\citep{x1,x2}\footnote{{\sc xillver} is a non-relativistic reflection model: it calculates the reflection spectrum generated by the illumination of a cold material (not necessarily a disk) by an incident radiation. Disk geometry and spacetime metric are included after convolving the output of {\sc xillver} with a convolution model. In the Kerr metric for an infinitesimally thin disk, this is done with {\sc relconv} and the result is {\sc relxill}: in XSPEC language, {\sc relxill} = {\sc relconv} $\times$ {\sc xillver}. In this work, we simply replace {\sc relconv} with a convolution model for thick disks.}. Once we have the theoretical reflection spectrum of a given configuration (specified by the parameters of the spacetime, of the disk, of {\sc xillver}, and by the viewing angle), we use XSPEC and the \verb9fakeit9 command to generate the simulated observations. For the X-ray mission, we choose \textsl{NICER}~\citep{nicer}, and therefore we use its background, ancillary, and response matrix files. \textsl{NICER} has a good energy resolution near the iron line, which make it suitable for X-ray reflection spectroscopy. It was launched in 2017 and there is already some important work in literature on spin measurements of black hole binaries using the iron line method with this mission.

In XSPEC language, the simulations are done with the model

\vspace{0.2cm}

{\sc tbabs $\times$ ( powerlw + reflection )} ,

\vspace{0.2cm}

\noindent where {\sc tbabs} takes the Galactic absorption into account~\citep{wilms}, {\sc powerlw} describes the continuum from the corona, and {\sc reflection} is the reflection component obtained by applying our convolution model to the output of {\sc xillver}.

First, we consider four simulations in the Kerr spacetime. The black hole spin parameter, $a_*$, is set to 0.8 (simulation~A), 0.85 (simulation~B), 0.9 (simulation~C), and 0.95 (simulation~D). For all the simulations, the inclination angle is fixed to $35^\circ$ and in {\sc xillver} we adopt the ionization $\log\xi = 3.1$ ($\xi$ in units erg~cm~s$^{-1}$), the iron abundance $A_{\rm Fe} = 1$ (Solar abundance), and the photon index $\Gamma = 2$. We set the outer edge of the accretion disk at $R_{\rm out} = 40$~$M$ for all simulations. The inner edges of the disk turns out to be $R_{\rm in} = 2.129$~$M$ (simulation~A), 1.951~$M$ (simulation~B), 1.750~$M$ (simulation~C), and 1.507~$M$ (simulation~D). The emissivity profile of the reflection spectrum is described by a power-law with emissivity index $q = 9$ in order to limit the effects of the outer edge of the disk. We set the photon flux around $1.4 \cdot 10^{-10}$~erg~cm$^{-2}$~s$^{-1}$ and the exposure time around 420~ks in order to have about 50~million photons in the energy range 0.2-12~keV. Such a photon count is higher than the typical photon count for an observation of a bright active galactic nucleus, so we can potentially better constrain the model parameters than what it is possible today with current X-ray observations.

The simulated data are analyzed with the XSPEC model

\vspace{0.2cm}

{\sc tbabs $\times$ relxill\_nk} ,

\vspace{0.2cm}

\noindent where {\sc relxill\_nk} is our relativistic reflection model~\citep{noi1,noi2}. We do not assume the Kerr spacetime and we employ the Johannsen metric with a possible non-vanishing deformation parameter $\alpha_{13}$~\citep{tj}. The Johannsen metric is briefly reviewed in Appendix~\ref{a-j}. It is not a solution of a well-defined theory of gravity but a parametric black hole metric in which an infinite number of deformation parameters are introduced to quantify possible deviations from the Kerr background. When all deformation parameters vanish, we exactly recover the Kerr solution. With such a metric, it is like performing a null experiment: we want to test the Kerr nature of a black hole and we try to measure the deformation parameters of the Johannsen metric to check whether they vanish.

\begin{figure*}
\vspace{0.3cm}
\includegraphics[width=15cm]{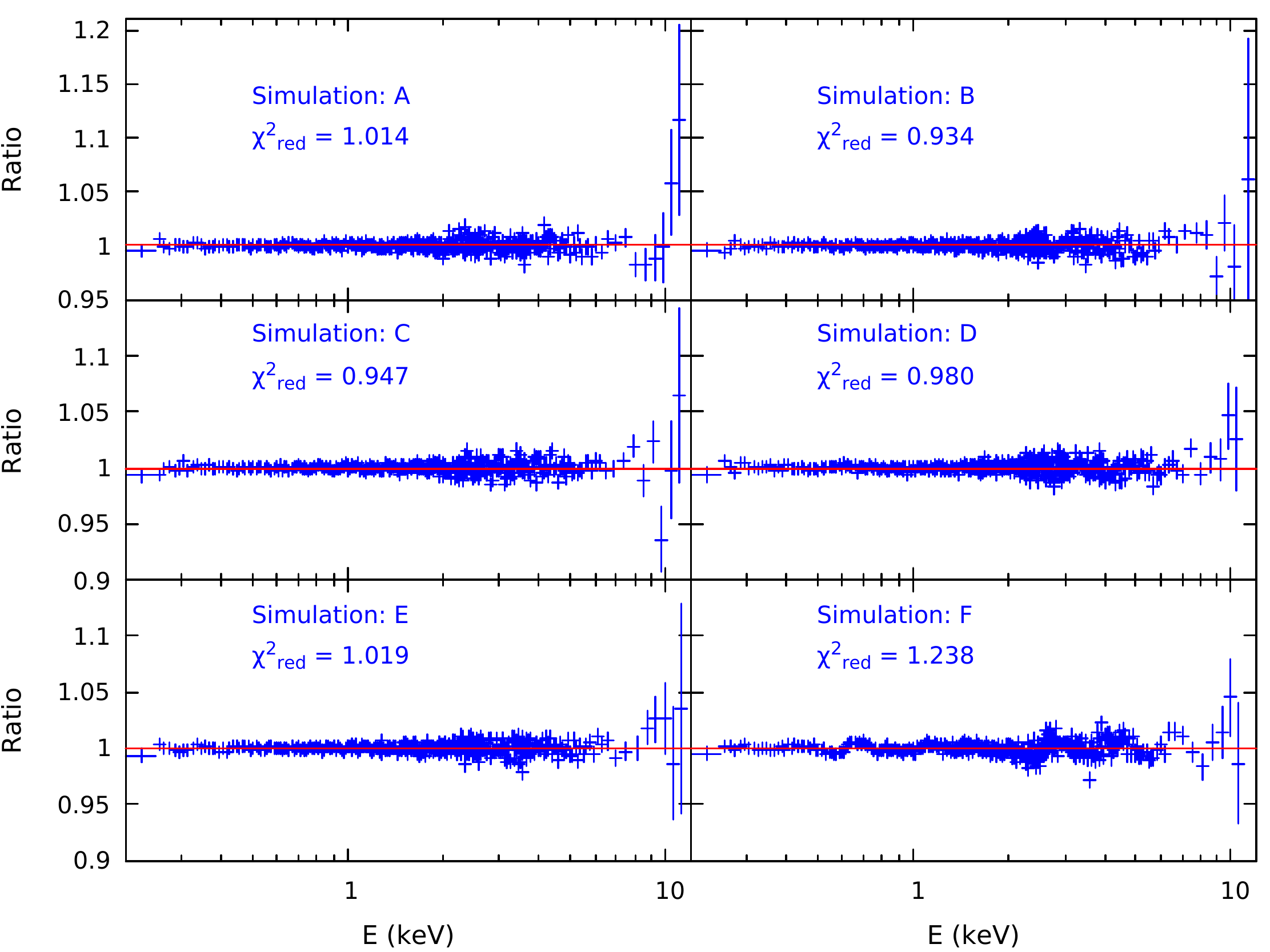}
\caption{Ratios between data and best-fit model for simulations A-F. Each panel also shows the reduced $\chi^2$ of the best-fit.}
\label{f-model_ratio}
\end{figure*}

Tab.~\ref{t-fit} shows the results of our analysis of simulations~A-D. The quality of the best-fit can be seen from Fig.~\ref{f-model_ratio}, which shows the ratios between the data and the best-fit models. The results of the analysis of these simulations are discussed in the next section.

We consider two more simulations imposing a non-Kerr background. The black hole spin parameter is set to 0.95 and the deformation parameter is $\alpha_{13} = 0.3$ (simulation~E) and $-0.3$ (simulation~F). The values of the other model parameters are the same as simulations~A-D. Even for simulations E and F we use \textsl{NICER} and we require about 50~million photons in the energy range 0.2-12~keV. We analyze the data with {\sc relxill\_nk} and we report best-fit values and ratio plots in Tab.~\ref{t-fit2} and in the two bottom panels of Fig.~\ref{f-model_ratio}, respectively. 

%%%%%%%%%%%%%%%%%%%%%%%%%%%%%%%%%%%%%%%%%%%%%%%%%%%%%%%%%%%%%%%%%%%%%%%%%%
\subsection {Slab reflector vs Toroidal reprocessor}

For the calculation of reflection spectra, we have assumed the Polish donut is optically thick through the use of a cold slab reflector. However, there can be scenarios in which the disk is not optically thick. For this reason, here we perform a brief comparison of the iron line generated by the optically thick model we have used throughout this work to that of a model that includes the effect of scattering that would be present in a disk that is not totally opaque. For the scattering model we use the Compton-thick toroidal reprocessor created by \citet{my2009} called MYTorus. The geometry used is that of a torus with a circular cross-section and is placed such that it subtends a solid angle of $2\pi$ at the central source, which emits isotropically. The model includes the effect of both the reflected and scattered emission from the cold (neutral) reprocessor (for more details see \citet{my2009}). Since the X-ray source used in MYTorus is a centrally-located point source, while in the model used throughout this work nothing is assumed about the location or geometry of the source the comparison here is not exact but rather an approximate look at the effect of including scattering.

\begin{figure*}
\vspace{0.3cm}
\includegraphics[width=17.5cm]{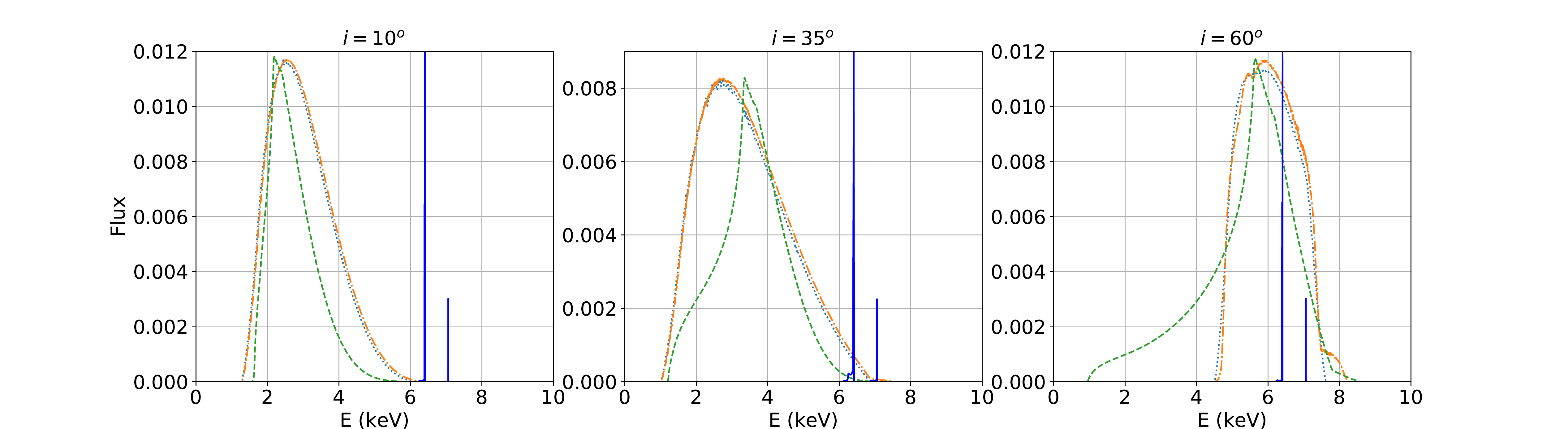}
\caption{Comparision of iron lines obtained from Polish donut disk (blue dotted curve), Polish donut $\times$ MYTorus iron line (orange dotted-dashed curve), {\sc relconv} $\times$ MYTorus iron line (green dashed curve) and the iron Fe K$\alpha$ and K$\beta$ emission lines obtained from MYTorus (blue solid lines). The inclination angle are chosen to be 10$^\circ$ (left panel), 35$^\circ$(middle panel) and 60$^\circ$(right panel). The values of parameters used to calculate the Polish donut iron lines are $a_* = 0.95$, disk outer edge $R_{\rm out} = 40$~$M$ and emissivity index $q = 9$.}
\label{pd_vs_myt}
\end{figure*}

For the comparison we will only study the effect on the iron line as the difference is much more noticeable there than in the full reflection spectrum. We compare the following four models in Fig.~\ref{pd_vs_myt}

\begin{enumerate}
\item{MYTorus iron line}
\item{Polish donut iron line}
\item{{\sc relconv} $\times$ MYTorus iron line}
\item{Polish donut $\times$ MYTorus iron line}
\end{enumerate}

\noindent where "iron line" means we only take the iron line part of the model and $\times$ refers to a convolution. The blue solid lines are the Fe K$\alpha$ and K$\beta$ emission lines produced by the model MYTorus. These lines are intrinsically narrow because of the fact that MYTorus model does not include relativistic effects. These lines contain both the zeroth-order photons as well as the scattered photons. The input parameters used to generate these lines are: column density $N_H = 2 \times 10^{24}$ cm$^{-2}$, photon index $\Gamma = 2$, redshift $z = 0$. We used the version of the fits table to generate the iron lines for which offset energy of the line is 0 and the termination energy of the incident power-law continuum is 300~keV. In order to introduce the relativistic effects in the MYTorus iron lines we convolved it with the convolution model {\sc relconv}. The parameters used for {\sc relconv} are: spin $a* = 0.95$, emissivity index $q = 9$ and outer edge of the disk $R_{\rm out} = 40$~$M$. In this case the relativistic effects are more important than the geometry of the disk. The resultant iron line becomes broadened (green dashed curve, Fig. \ref{pd_vs_myt}). The effects of convolving the iron lines produced by MYTorus with the Polish donut model are shown by the orange dotted-dashed curves. In this case the geometry of the disk becomes important and the iron lines show significant differences as compared to the {\sc relconv}$\times$ MYTorus iron line (green dashed curves). However, the differences are negligible when the Polish donut $\times$ MYTorus iron line is compared with the iron lines produced by the Polish donut model. Thus, it seems that including scattering has some effect on the iron line, and in turn the full reflection spectrum, however this effect is likely not large enough to have a significant impact on recovered parameters.

%%%%%%%%%%%%%%%%%%%%%%%%%%%%%%%%%%%%%%%%%%%%%%%%%%%%%%%%%%%%%%%%%%%%%%%%

\section{Discussion and conclusions}
\label{s-con}

In the previous section, we have simulated observations of reflection spectra from thick accretion disks in Kerr (simulations~A-D) and non-Kerr (simulations E and F) spacetimes and we have analyzed the data with the thin disk relativistic reflection model {\sc relxill\_nk} to figure out potential systematic uncertainties when we fit the reflection spectrum of a source accreting at Eddington or super-Eddington rate with a thin disk model. Note that this is the case shown in many studies in the literature of supermassive black holes. The Polish donut disk employed in our simulations is quite a simple model far from a real accretion disk, but it is used here to catch the basic differences between thin and thick disks and the analytic model helps to simplify our study.

Let us first discuss the results of simulations~A-D. From Tab.~\ref{t-fit} and the ratio plots in Fig.~\ref{f-model_ratio}, we can say:
\begin{enumerate}
\item The quality of the fits is good: the reduced $\chi^2$ is close to 1 and we do not see unresolved features in the ratio plots. In other words, reflection spectra from thick disks can look like those from thin disks. This may sound a little bit surprising at first, because from Figs.~\ref{f-lines3} and \ref{f-lines9} we see that the iron line shapes of thick disks are very different from the same lines generated by thin disks. However, in Figs.~\ref{f-lines3} and \ref{f-lines9} all the other model parameters are fixed and we only change the disk structure. In reality, we have several free parameters and, eventually, reflection spectra from thick disks can look like those from thin disks with different model parameters.  
\item The fit can recover correct values of some model parameters and wrong values of other model parameters. For example, the fit can measure more or less the right value of $\log\xi$, $A_{\rm Fe}$, and $\Gamma$ even if we employ a thin disk model to fit the spectrum of a thick disk. On the contrary, the estimates of the emissivity index $q$, the spin parameter $a_*$, and the deformation parameter $\alpha_{13}$ are strongly affected by the systematic uncertainties related to the use of a wrong model.
\item The spin parameter $a_*$ is always and significantly overestimated. This result could be expected, because the inner edge of the disk is inside the ISCO in all the simulations and therefore some photons are strongly redshifted, as it would be the case of emission from the ISCO in a spacetime with smaller ISCO radius.
\item While the simulations are done in a Kerr spacetime, the fit ``typically'' measures a non-vanishing value of the deformation parameter $\alpha_{13}$. Generally speaking, this is also a result that could be expected: we are fitting the data with a wrong model and therefore we find a wrong estimate of the deformation parameter. Interestingly, the best-fit of $\alpha_{13}$ is far from the Kerr solution for simulation~A and moves to the Kerr solution $\alpha_{13} = 0$ as the spin parameter increases, to recover the Kerr metric in simulation~D (but with too large a value of the spin parameter with respect to its input value in the simulation). We are tempted to argue that, when the inner edge of the disk is very close to the black hole, relativistic effects in the reflection spectrum are very important so that even small deviations from Kerr could produce very different reflection spectra. Note also that the results of simulation~D seem to nicely explain the very strong constraints on the Kerr hypothesis found in \citet{a1}, where we analyzed some bare active galactic nuclei: some sources look like near-extremal Kerr black holes, with the best-fit of the spin parameter $a_*$ stuck at the maximum value of {\sc relxill\_nk}, namely 0.998, and very good constraints on the deformation parameter $\alpha_{13}$. This is exactly what we find in simulation~D. If such an interpretation is correct, we should be very skeptical of the extremely high values of the spin parameters found in \citet{a1} for those sources.
\end{enumerate}

Simulations E and F in non-Kerr spacetime mainly confirm the conclusions derived from the results of simulations~A-D. The fit can recover the correct value of some parameters ($\log\xi$, $A_{\rm Fe}$, and $\Gamma$) and infers wrong values for other parameters ($q$, $a_*$, and $\alpha_{13}$). For simulations~F, in which the input value is $\alpha_{13} = -0.3$, the quality of the fit is not good: the reduced $\chi^2$ is around 1.2 and its ratio plot shows some unresolved features. In all simulations, in Kerr and non-Kerr spacetimes but with the exception of simulation~D, the best-fit value of $\alpha_{13}$ is negative. This, presumably, is just a property of the deformation parameter $\alpha_{13}$, which can somehow mimic the effect of a Polish donut disk on the reflection spectrum when it becomes negative. For simulation~D, the inner edge of the disk is probably so close to the black hole event horizon that relativistic effects are more important than the disk structure. This would explain the trend shown in Tab.~\ref{t-fit}, where the estimate of $\alpha_{13}$ moves to the Kerr case $\alpha_{13} = 0$ as the inner edge of the disk gets closer to the black hole event horizon.

In conclusion, the results of our work suggest that black hole spin measurements and tests of the Kerr hypothesis from the study of relativistically broadened reflection features can be significantly affected by systematic uncertainties when we employ a thin disk model to analyze sources accreting at the Eddington or super-Eddington rate, as is often the case for many supermassive black holes.

%%%%%%%%%%%%%%%%%%%%%%%%%%%%%%%%%%%%%%%%%%%%%%%%%%

\section*{Acknowledgements}

This work was supported by the Innovation Program of the Shanghai Municipal Education Commission, Grant No.~2019-01-07-00-07-E00035, and Fudan University, Grant No.~IDH1512060. S.N. acknowledges support from the Alexander von Humboldt Foundation and the Excellence Initiative at Eberhard-Karls Universit\"at T\"ubingen.

%%%%%%%%%%%%%%%%%%%%%%%%%%%%%%%%%%%%%%%%%%%%%%%%%%



%\bibliographystyle{mnras}

\begin{thebibliography}{99}

\bibitem[Abdikamalov et al.(2019a)]{noi2} Abdikamalov, A.~B., Ayzenberg, D., Bambi, C., et al.\ 2019a, \apj, 878, 91

\bibitem[Abdikamalov et al.(2019b)]{t5} Abdikamalov, A.~B., Ayzenberg, D., Bambi, C., et al.\ 2019b, arXiv e-prints, arXiv:1905.08012

\bibitem[Abramowicz et al.(1978)]{abramowicz1978} Abramowicz, M., Jaroszynski, M., \& Sikora, M.\ 1978, \aap, 63, 221

\bibitem[Bambi(2012)]{cfm} Bambi, C.\ 2012, \apj, 761, 174

\bibitem[Bambi(2013)]{t3} Bambi, C.\ 2013, \prd, 87, 023007  

\bibitem[Bambi(2017a)]{review} Bambi, C.\ 2017a, Reviews of Modern Physics, 89, 025001

\bibitem[Bambi(2017b)]{book} Bambi, C.\ 2017b, {\it Black Holes: A Laboratory for Testing Strong Gravity} (Springer Singapore),  doi:10.1007/978-981-10-4524-0

\bibitem[Bambi(2018)]{i7} Bambi, C.\ 2018, Annalen der Physik, 530, 1700430

\bibitem[Bambi et al.(2017)]{noi1} Bambi, C., C{\'a}rdenas-Avenda{\~n}o, A., Dauser, T., Garc{\'{\i}}a, J.~A., \& Nampalliwar, S.\ 2017, \apj, 842, 76

\bibitem[Brenneman(2013)]{t1} Brenneman, L.\ 2013, Measuring the Angular Momentum of Supermassive Black Holes, SpringerBriefs in Astronomy~ISBN 978-1-4614-7770-9 (Springer, New York, New York)

\bibitem[Brenneman \& Reynolds(2006)]{kc} Brenneman, L.~W., \& Reynolds, C.~S.\ 2006, \apj, 652, 1028

\bibitem[Cao et al.(2018)]{t4} Cao, Z., Nampalliwar, S., Bambi, C., et al.\ 2018, \prl, 120, 051101

\bibitem[Dauser et al.(2013)]{rc2} Dauser, T., Garcia, J., Wilms, J., et al.\ 2013, \mnras, 430, 1694

\bibitem[Dauser et al.(2010)]{rc1} Dauser, T., Wilms, J., Reynolds, C.~S., \& Brenneman, L.~W.\ 2010, \mnras, 409, 1534 

\bibitem[Fabian et al.(2000)]{i3} Fabian, A.~C., Iwasawa, K., Reynolds, C.~S., \& Young, A.~J.\ 2000, \pasp, 112, 1145

\bibitem[Fabian et al.(1989)]{i1} Fabian, A.~C., Rees, M.~J., Stella, L., \& White, N.~E.\ 1989, \mnras, 238, 729

\bibitem[Fishbone \& Moncrief(1976)]{FM1976} Fishbone, L.~G., \& Moncrief, V.\ 1976, \apj, 207, 962

\bibitem[Garc{\'{\i}}a et al.(2013)]{x2} Garc{\'{\i}}a, J., Dauser, T., Reynolds, C.~S., et al.\ 2013, \apj, 768, 146

\bibitem[Garc{\'{\i}}a \& Kallman(2010)]{x1} Garc{\'{\i}}a, J., \& Kallman, T.~R.\ 2010, \apj, 718, 695

\bibitem[Gendreau et al.(2016)]{nicer} Gendreau, K.~C., Arzoumanian, Z., Adkins, P.~W., et al.\ 2016, \procspie, 99051H

\bibitem[Johannsen(2013)]{tj} Johannsen, T.\ 2013, \prd, 88, 044002

\bibitem[Kozlowski et al.(1978)]{kja78} Kozlowski, M., Jaroszynski, M., \& Abramowicz, M.\ 1978, \aap, 63, 209

\bibitem[Li \& Bambi(2013)]{zilong} Li, Z., \& Bambi, c.\ 2013, \jcap, 03, 031

\bibitem[Liu et al.(2019)]{honghui} Liu, H., Abdikamalov, A.~B., Ayzenberg, D., et al.\ 2019, \prd, 99, 123007

\bibitem[McClintock et al.(2014)]{isco2} McClintock, J.~E., Narayan, R., \& Steiner, J.~F.\ 2014, \ssr, 183, 295

\bibitem[Reynolds(2014)]{t2} Reynolds, C.~S.\ 2014, \ssr, 183, 277

\bibitem[Risaliti et al.(2013)]{i6} Risaliti, G., Harrison, F.~A., Madsen, K.~K., et al.\ 2013, \nat, 494, 449

\bibitem[Ross \& Fabian(2005)]{rl} Ross, R.~R., \& Fabian, A.~C.\ 2005, \mnras, 358, 211

\bibitem[Steiner et al.(2010)]{isco1} Steiner, J.~F., McClintock, J.~E., Remillard, R.~A., et al.\ 2010, \apjl, 718, L117

\bibitem[Tanaka et al.(1995)]{i2} Tanaka, Y., Nandra, K., Fabian, A.~C., et al.\ 1995, \nat, 375, 659

\bibitem[Taylor \& Reynolds(2018)]{tr18} Taylor, C., \& Reynolds, C.~S.\ 2018, \apj, 855, 120

\bibitem[Tripathi et al.(2019b)]{a2} Tripathi, A., Nampalliwar, S., Abdikamalov, A.~B., et al.\ 2019b, \apj, 875, 56

\bibitem[Tripathi et al.(2018)]{a0} Tripathi, A., Nampalliwar, S., Abdikamalov, A.~B., et al.\ 2018, \prd, 98, 023018

\bibitem[Tripathi et al.(2019a)]{a1} Tripathi, A., Yan, J., Yang, Y., et al.\ 2019a, \apj, 874, 135

\bibitem[Wald(1984)]{wald} Wald, R.~M.\ 1984, {\it General Relativity} (University of Chicago Press, Chicago, Illinois)

\bibitem[Walton et al.(2012)]{i5} Walton, D.~J., Reis, R.~C., Cackett, E.~M., Fabian, A.~C., \& Miller, J.~M.\ 2012, \mnras, 422, 2510

\bibitem[Walton et al.(2013)]{w13} Walton, D.~J., Nardini, E., Fabian, A.~C., Gallo, L.~C., \& Reis, R.~C.\ 2013, \mnras, 428, 2901

\bibitem[Wilms et al.(2000)]{wilms} Wilms, J., Allen, A., \& McCray, R.\ 2000, \apj, 542, 914

\bibitem[Wu \& Wang(2007)]{ww07} Wu, S.-M., \& Wang, T.-G.\ 2007, \mnras, 378, 841 

\bibitem[Xu et al.(2018)]{yerong} Xu, Y., Nampalliwar, S., Abdikamalov, A.~B., et al.\ 2018, \apj, 865, 134

\bibitem[Zhang et al.(2019)]{yuexin} Zhang, Y., Abdikamalov, A.~B., Ayzenberg, D., et al.\ 2019, \apj, 875, 41

\bibitem[Zoghbi et al.(2010)]{i4} Zoghbi, A., Fabian, A., Uttley, P., Miniutti, G., Gallo, L., Reynolds, C., Miller, J., \& Ponti, G.\ 2010, \mnras, 401, 2419

\bibitem[Murphy \& Yaqoob(2009)]{my2009}Murphy K.~D.,Yaqoob T.,2009,\mnras, 397,1549

\end{thebibliography}


%%%%%%%%%%%%%%%%%%%%%%%%%%%%%%%%%%%%%%%%%%%%%%%%%%

%%%%%%%%%%%%%%%%% APPENDICES %%%%%%%%%%%%%%%%%%%%%

\appendix

\section{Johannsen metric}
\label{a-j}

The Johannsen background is a parametric black hole spacetime designed to quantify possible deviations from the Kerr solution~\citep{tj}. In Boyer-Lindquist-like coordinates, its line element reads
\begin{eqnarray}
ds^2 &=&-\frac{\tilde{\Sigma}\left(\Delta-a^2A_2^2\sin^2\theta\right)}{B^2}dt^2 
+\frac{\tilde{\Sigma}}{\Delta}dr^2+\tilde{\Sigma} d\theta^2 \nonumber\\
&&-\frac{2a\left[\left(r^2+a^2\right)A_1A_2-\Delta\right]\tilde{\Sigma}\sin^2\theta}{B^2}dtd\phi \nonumber\\
&&+\frac{\left[\left(r^2+a^2\right)^2A_1^2-a^2\Delta\sin^2\theta\right]\tilde{\Sigma}\sin^2\theta}{B^2}d\phi^2 \, , \qquad
\end{eqnarray}
where $M$ is the black hole mass, $a = J/M$, $J$ is the black hole spin angular momentum, $\tilde{\Sigma} = \Sigma = f$, and
\begin{eqnarray}
\Sigma &=& r^2 + a^2 \cos^2\theta \, , \\
\Delta &=& r^2 - 2 M r + a^2 \, , \\
B &=& \left(r^2+a^2\right)A_1-a^2A_2\sin^2\theta \, .
\end{eqnarray}
The functions $A_1$, $A_2$, $A_5$, and $f$ are defined as
\begin{eqnarray}
A_1 &=& 1 + \sum^\infty_{n=3} \alpha_{1n} \left(\frac{M}{r}\right)^n \, , \\
A_2 &=& 1 + \sum^\infty_{n=2} \alpha_{2n}\left(\frac{M}{r}\right)^n \, , \\
A_5 &=& 1 + \sum^\infty_{n=2} \alpha_{5n}\left(\frac{M}{r}\right)^n \, , \\
f &=& \sum^\infty_{n=3} \epsilon_n \frac{M^n}{r^{n-2}} \, ,
\end{eqnarray}
and $\{ \alpha_{1n} \}$, $\{ \alpha_{2n} \}$, $\{ \alpha_{5n} \}$, and $\{ \epsilon_n \}$ are four sets of deformation parameters without constraints from the Newtonian limit and experiments in the Solar System. In the present manuscript, as a preliminary study on the impact of the disk structure on tests of the Kerr hypothesis, we have only considered the possibility of a non-vanishing $\alpha_{13}$ and set to zero all other deformation parameters.

%%%%%%%%%%%%%%%%%%%%%%%%%%%%%%%%%%%%%%%%%%%%%%%%%%

% Don't change these lines
\bsp	% typesetting comment
\label{lastpage}
\end{document}